\documentclass[notitlepage,amsmath,preprintnumbers,nofootinbib,aps,11pt]{revtex4-2} 

\pdfoutput=1

\usepackage{amsmath,amssymb,url}
\usepackage{amssymb,amsfonts,multirow}
\usepackage{rotating}
\usepackage{color}
\usepackage{hhline}
\usepackage{slashed}

\usepackage{xcolor}
\usepackage{multirow}

\usepackage{array,multirow}

\usepackage{booktabs}

\definecolor{Gray}{gray}{0.85}
\definecolor{LightGreen}{rgb}{0.88, 1, 0.88}
\definecolor{Lime}{rgb}{0,255,0}
\definecolor{LightCyan}{rgb}{0.88,1,1}
\definecolor{LightRed}{rgb}{1, 0.85, 0.85}
\definecolor{Red}{rgb}{1, 0, 0}
\definecolor{LightYellow}{rgb}{1, 1, 0.85}
\definecolor{Yellow}{rgb}{1,1,0.05}
\definecolor{LightBlue}{rgb}{0.87, 0.94, 1}
\definecolor{white}{gray}{1}
\definecolor{black}{gray}{0}
\newcolumntype{G}{>{\columncolor{LightGray}}c}

\newcolumntype{?}{!{\vrule width 1pt}}
\newcolumntype{`}{!{\vrule width 1.5pt}}

\newcommand{\gam}[2]{\gamma_{#1}^{(#2)}}

\usepackage{framed}
\usepackage{bm}
\usepackage{amsmath}
\usepackage{graphicx}
\usepackage{amssymb}
\usepackage{epstopdf}
\usepackage{hyperref}
\usepackage{subfigure}
\usepackage{epstopdf}
\usepackage{verbatim} 
\usepackage{array}
\usepackage{booktabs}
\usepackage{color}

\def\beq{\begin{equation}}
\def\eeq{\end{equation}}

\def\bea{\arraycolsep .1em \begin{eqnarray}}
\def\eea{\end{eqnarray}}
\def\Tr{{\rm Tr}}
\newcommand{\step}{\vspace{.5em}}

\def\eps{\epsilon}

\def\eq#1{(\ref{#1})}

\def\s0#1#2{\mbox{\small{$ \frac{#1}{#2} $}}}
\def\0#1#2{\frac{#1}{#2}}

\def\grgl{\:\hbox to -0.2pt{\lower2.5pt\hbox{$\sim$}\hss}{\raise3pt\hbox{$>$}}\:}
\def\klgl{\:\hbox to -0.2pt{\lower2.5pt\hbox{$\sim$}\hss}{\raise3pt\hbox{$<$}}\:}

\newcommand \be {\begin{equation}}
\newcommand \ee {\end{equation}}
\newcommand \bed {\begin{displaymath}}
\newcommand \eed {\end{displaymath}}

\newcommand{\bit}{\begin{itemize}}
\newcommand{\eit}{\end{itemize}}

\usepackage{colortbl}

\definecolor{Gray}{gray}{0.85}
\definecolor{LightGray}{gray}{0.93}
\definecolor{LightGreen}{rgb}{0.88, 1, 0.88}
\definecolor{LightCyan}{rgb}{0.88,1,1}
\definecolor{LightRed}{rgb}{1, 0.85, 0.85}
\definecolor{LightRed}{rgb}{1, 0.88, 0.88}
\definecolor{LightYellow}{rgb}{1, 1, 0.85}
\definecolor{LightBlue}{rgb}{0.87, 0.94, 1}
\definecolor{white}{gray}{1}

\usepackage{array,mathtools,amssymb,booktabs}
\newcolumntype{C}{>{$}c<{$}}
\AtBeginDocument{
\heavyrulewidth=.16em
\lightrulewidth=.1em
\cmidrulewidth=.03em
\belowrulesep=.4ex
\belowbottomsep=0pt
\aboverulesep=.4ex
\abovetopsep=0pt
\cmidrulesep=\doublerulesep
\cmidrulekern=.5em
\defaultaddspace=.5em
}

\makeatletter
    \def\CT@@do@color{%
      \global\let\CT@do@color\relax
            \@tempdima\wd\z@
            \advance\@tempdima\@tempdimb
            \advance\@tempdima\@tempdimc
    \advance\@tempdimb\tabcolsep
    \advance\@tempdimc\tabcolsep
    \advance\@tempdima2\tabcolsep
            \kern-\@tempdimb
            \leaders\vrule
                    \hskip\@tempdima\@plus  1fill
            \kern-\@tempdimc
            \hskip-\wd\z@ \@plus -1fill }
            
\begin{document}

\title{Asymptotic safety  guaranteed for strongly coupled gauge theories}
\author{Andrew D.~Bond}

\author{Daniel F.~Litim}

\affiliation{\mbox{Department of Physics and Astronomy, U Sussex, Brighton, BN1 9QH, U.K.}}

\begin{abstract}
 We demonstrate that  interacting ultraviolet fixed points in four dimensions exist at strong coupling, and away from large-$N$ Veneziano limits.
  This is established   exemplarily for    semi-simple  supersymmetric gauge theories with chiral matter and  superpotential interactions     by using the renormalisation group  and exact methods from supersymmetry.
 We determine the   entire superconformal  window of ultraviolet fixed points  as a function of   field multiplicities.
 Results are in accord with  the $a$-theorem, bounds on conformal charges, Seiberg duality, and unitary. 
 We also find  manifolds of Leigh-Strassler models exhibiting lines of infrared fixed points. At weak coupling, findings  are    confirmed  using   perturbation theory up to three loop.   Benchmark models with low field multiplicities are provided  including examples with Standard~Model-like gauge sectors.
 Implications for particle physics, model building, and conformal field theory are indicated.

\end{abstract}

\maketitle

\section{Introduction}

Ultraviolet fixed points are key for the renormalisability and  predictive power of quantum field theories. 
 The classic example is given by asymptotic freedom of the strong nuclear force, where the 
fixed point is non-interacting   \cite{Gross:1973id, Politzer:1973fx}. The possibility of interacting ultraviolet (UV) fixed points, often denoted   as asymptotic safety \cite{Weinberg:1980gg},
has  been conjectured  early on \cite{Bailin:1974bq}.  
Recently, this field  has taken up some speed due to the  discovery  of UV conformal fixed points   in models of particle physics 
\cite{Litim:2014uca,Bond:2018oco,Bond:2016dvk,Bond:2017sem,Bond:2019npq,Bond:2017lnq}. 
Conditions under which asymptotic safety arises in weakly coupled four-dimensional quantum field theories (without gravity) 
are by now well understood: Non-abelian gauge fields are central  \cite{Bond:2018oco}, alongside Yukawa and scalar 
interactions and subject to a stable vacuum \cite{Bond:2016dvk,Bond:2017sem}. Templates
with strict perturbative control have been found for unitary \cite{Litim:2014uca},
orthogonal and symplectic \cite{Bond:2019npq}, or  product gauge groups  \cite{Bond:2017lnq}, and supersymmetry \cite{Bond:2017suy}. This has also triggered  new ideas for  model building \cite{Bond:2017wut,Kowalska:2017fzw}, asymptotically safe extensions of the Standard Model explaining the electron and muon $g-2$ anomalies \cite{Hiller:2019tvg,Hiller:2019mou,Hiller:2020fbu} whose new type of flavor phenomenology can be tested at colliders  \cite{Bissmann:2020lge}, and explanations of flavour anomalies as evidenced in rare $B$-meson decays~\cite{Bause:2021prv}.
Further results cover  vacuum stability  \cite{Litim:2015iea} including the Higgs \cite{Hiller:2019tvg,Hiller:2019mou,Hiller:2020fbu,Bause:2021prv}, abelian factors \cite{Kowalska:2017fzw},  global fixed points \cite{Buyukbese:2017ehm}, 
aspects of radiative symmetry breaking \cite{Abel:2017ujy}, UV conformal  windows  \cite{Bond:2017tbw}, and fixed point mergers \cite{Bond:2021tgu}.
 
Despite the vast body of weakly coupled fixed points at hand, it has remained an open challenge to understand  asymptotic safety of strongly coupled 4d quantum field theories from first principles. It would be desirable to have rigorous and explicit  examples at hand, if only as a proof of principle, and to clarify  whether large matter field anomalous dimensions  or new phenomena may become an obstacle for  safety in the UV. 
Moreover, it is well known that weakly coupled fixed points often require a large number of gauge and matter fields, such as in a Veneziano large-$N$ limit,  while decreasing the number of matter fields turns fixed point interactions  stronger. 
In the context of asymptotically safe model building where only finitely many new matter fields are added to the Standard Model \cite{Bond:2017wut,Kowalska:2017fzw,Hiller:2019tvg,Hiller:2019mou,Hiller:2019mou,Hiller:2020fbu,Bissmann:2020lge,Bause:2021prv}, it  becomes paramount to control the size  of UV conformal windows non-perturbatively, and to understand how few matter fields can sustain an underlying fixed point  \cite{Bond:2017tbw,Bond:2021tgu}.
It would be equally important to  understand whether or not qualitatively new types of fixed points  arise at strong coupling, beyond those discovered at weak coupling.

In this spirit, we put forward a non-perturbative search for interacting UV fixed points in conventional 4d quantum field theories, without gravity.
With asymptotically safe fixed points  being presently out of reach for lattice simulations or the conformal bootstrap, we turn, instead,  to $N=1$ global supersymmetry as the non-perturbative tool of choice:
Non-renormalisation theorems ensure that superpotential (Yukawa) couplings are renormalised non-perturbatively via the chiral superfield anomalous dimensions, and exact infinite order perturbative RG equations are available for gauge couplings \cite{Novikov:1983uc,Novikov:1985rd}. Further, at superconformal fixed points, anomalous dimensions of all chiral superfields can be determined unambigiously using the  method of $a$-maximisation \cite{Intriligator:2003jj}. 
Finally, independent quartics do not  arise and vacuum stability is  automatically guaranteed provided   gauge and Yukawas  take viable fixed points. 
 Taken together, these ingredients  prove sufficient  to determine  interacting  fixed points reliably, including at strong coupling.

\begin{figure}
\begin{center}
\vskip-0.2cm
\includegraphics[width=0.45\textwidth]{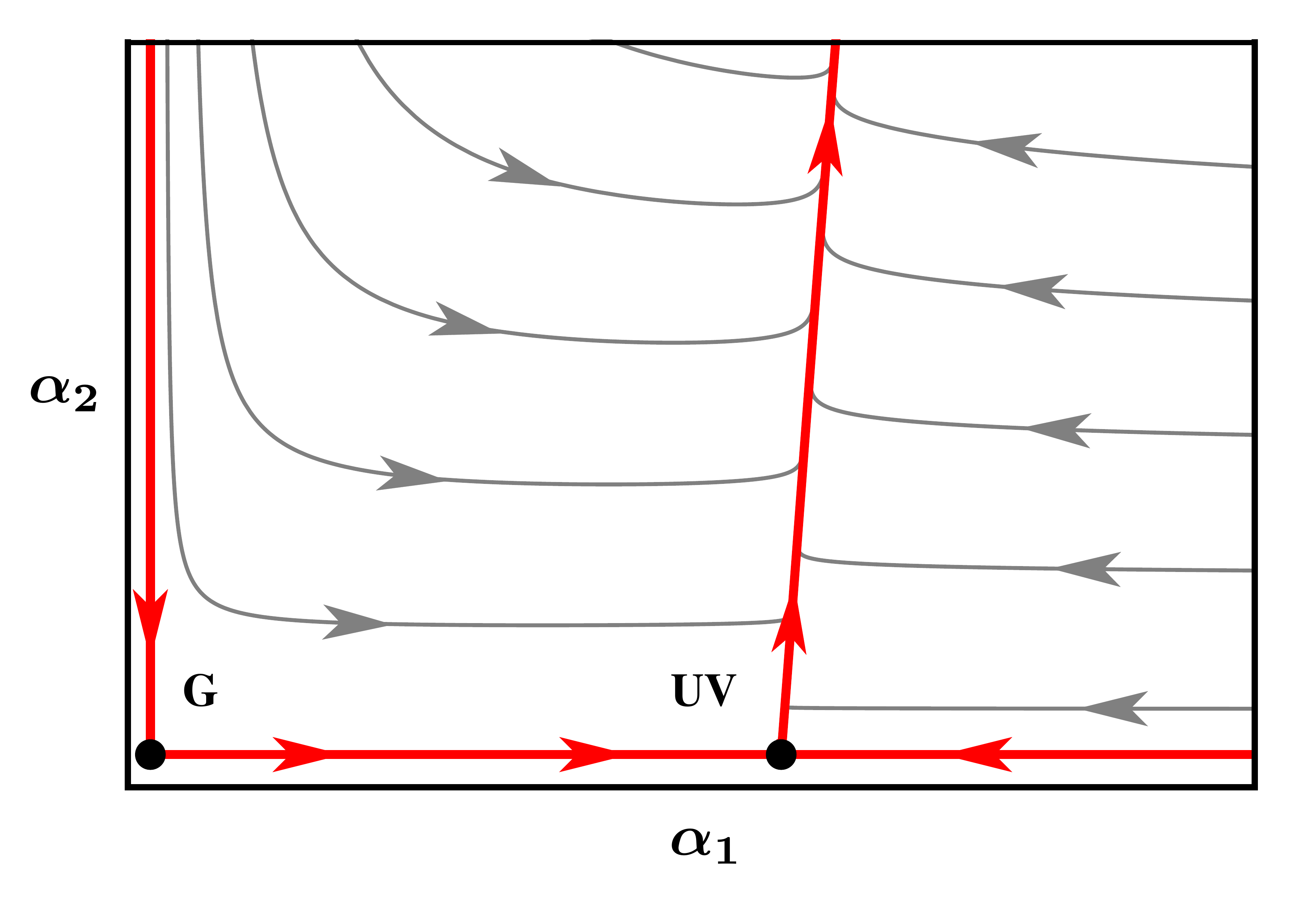}
\vskip-0.3cm
\caption{
The basic setup  in the plane of gauge couplings $(\alpha_1,\alpha_2)$, showing the free Gaussian fixed point (G) and an interacting ultraviolet fixed point (UV), with arrows on RG trajectories pointing from the UV to the IR. Note that the Gaussian is a ``saddle" and asymptotic freedom is absent. \label{pSchematic}
}
\end{center}
\vskip-.6cm 
\end{figure}

  Here, we apply this methodology exemplarily to an asymptotically non-free $SU(N_1)\times SU(N_2)$ supersymmetric gauge theory with massless chiral matter and a superpotential, and determine the entire conformal window of interacting UV fixed points. Our basic setup is illustrated in Fig.~\ref{pSchematic} which shows the phase diagram of a semi-simple gauge theory with superpotential interactions.  Notice that asymptotic freedom is absent because one of the gauge sector $(\alpha_2)$ is infrared free while the other not $(\alpha_1)$. The theory may nevertheless develop an asymptotically safe fixed point (UV) which allows as well-defined high energy limit and outgoing RG trajectories towards the IR, and which is the central topic of this study. Consistency with Seiberg duality, and constraints from the $a$-theorem, global charges, and  unitarity, are also observed  \cite{Mack:1975je}. Findings are confirmed independently using perturbation theory up to three loop order. Implications of our findings for the asymptotic safety conjecture, model building, and conformal field theory are indicated.

\begin{table}[b]
 \aboverulesep = 0mm
\belowrulesep = 0mm
\begin{center}
\begin{tabular}{`cc cc cc cc c`}
\toprule
\rowcolor{Yellow}
\ \ 
\bf Matter
&$\bm{\psi_L}$ 
&$\bm{\psi_R}$ 
&$\bm{\Psi_L}$ 
&$\bm{\Psi_R}$ 
&$\bm{\chi_L}$ 
&$\bm{\chi_R}$ 
&$\bm{Q_L}$ 
&$\bm{Q_R}$ \ \ 
\\
\midrule
&& && && && \\[-3mm]
$\ \ \bm{SU(N_1)}\ \ $
&$\overline{\Box}$
&$
\Box$
&$\Box$
&$\overline{\Box}$
&1
&1
&1
&1
\\
&& && && && \\[-3mm]
\rowcolor{LightGray}
&& && && && \\[-3mm]
\rowcolor{LightGray}
$\bm{SU(N_2)}$
&1
&1
&$\Box$
&$\overline{\Box}$
&$\overline{\Box}$
&$\Box$
&$\overline{\Box}$
&$\Box$
\\
&& && && && \\[-3mm]
&& && && && \\[-3mm]
{\bf Flavour}
&$N_F$
&$N_F$
&1
&1
&$N_F$
&$N_F$
&$N_Q$
&$N_Q$
\\[.4mm]
\bottomrule
\end{tabular}
\end{center}
 \vskip-.4cm
 \caption{\label{matter}Chiral matter with gauge charges and multiplicities.}
\end{table}

\section{Semi-Simple Gauge Theories with Matter}

We consider    semi-simple Yang-Mills theories  
with  product gauge group $SU(N_1)\times SU(N_2)$  coupled to
chiral superfields $(\psi,\chi,\Psi,Q)$ with flavour multiplicities $(N_F,N_F,1,N_Q)$ and gauge charges  as in Tab.~\ref{matter}.
We require that the model has a global $N=1$ supersymmetry. It is then characterised by two gauge  couplings $g_1$ and $g_2$ and a Yukawa coupling $y$ via the  superpotential
\beq\label{W}
W= y\,\Tr\big[\psi_L\,\Psi_L\,\chi_L+\psi_R\,\Psi_R\,\chi_R\big]\,,
\eeq
where the trace sums over flavour and gauge indices. 
The superfields $Q$ are not furnished with Yukawa interactions.
The theory has a  global
$SU(N_F)_L\times SU(N_F)_R\times SU(N_Q)_L\times SU(N_Q)_R$   flavour and a $U(1)_R$ symmetry and is characterised by the field multiplicities
\beq\label{Ns}
(N_1\,, \,N_2\,,\,  N_F\,,\,  N_Q)\,.
\eeq 
Asymptotic freedom and interacting infrared fixed points arise for suitable matter field multiplicities.
For this study, the regime of interest is where
the $SU(N_1)$ gauge sector is asymptotically free while the $SU(N_2)$ gauge sector is infrared free.
Accordingly,
the free theory 
corresponds to a ``saddle'' and asymptotic freedom cannot be achieved (Fig.~\ref{pSchematic}), very much like in the non-abelian gauge sectors 
of the Minimal Supersymmetric Standard Model (MSSM).  
 In this light, the gauge coupling $\alpha_2$ can be viewed as ``dangerously irrelevant'', in that it may to become relevant due to the gauge-Yukawa fixed point in the other gauge sector.

In the large $N$ Veneziano limit, the theory has previously been studied in perturbation theory, where it was found that interacting UV fixed points can arise at weak coupling with perturbatively small anomalous dimensions  \cite{Bond:2017suy}. 
The main point of this study is to investigate the conditions under which theories with \eq{Bs} may develop strongly interacting ultraviolet fixed points where anomalous dimensions become large, of order unity. 
 Ordinarily, this would require strong coupling methods such as lattice simulations to establish the claim. In supersymmetry, however,  powerful continuum methods  beyond perturbation theory  are available to which we turn next.\step\step

\section{Renormalisation Group}

To achieve our claim, we must find the  renormalisation group equations for all couplings and identify their fixed points, and non-perturbative expressions for the chiral superfield anomalous dimensions. Here, we exploit  key features   of $N=1$ supersymmetic gauge theories which make this task feasible. 

For supersymmetric gauge theories, closed  all-order expressions for perturbative $\beta$-functions   have been achieved by Novikov, Shifman, Vainstein, and Zakharov  (NSVZ) \cite{Novikov:1983uc,Novikov:1985rd} (see also \cite{Stepanyantz:2020uke,Korneev:2021zdz}). 
We introduce the gauge and Yukawa couplings as 
\beq
\alpha_{1,2}=
\left(\0{g_{1,2}}{4\pi}\right)^2\,,\quad\quad 
\alpha_y=
\left(\0{y}{4\pi}\right)^2\,,
\eeq  
and denote the chiral superfield anomalous dimensions   as $\gamma_a$   $(a=\psi,\Psi,\chi,Q)$,
with $\beta$-functions
$\beta_i\equiv d\alpha_i/d\ln \mu$ $(i=1,2,y)$   defined as usual.
The NSVZ beta functions for the gauge couplings of our models are then given by
\bea\label{beta1}
\beta_1&=&\frac{2\alpha_1^2}{F(\alpha_1)}[N_F\left(1-2\gamma_\psi\right)+N_2\left(1-2\gamma_\Psi\right)-3N_1]
\\
\label{beta2}
\beta_2&=&\frac{2\alpha_2^2}{F(\alpha_2)}[N_F\left(1-2\gamma_\chi\right)
+N_1\left(1-2\gamma_\Psi\right)
+N_Q\left(1-2\gamma_Q\right)-3N_2]\,.
\eea
The scheme-dependent function $F(\alpha)$ is normalised to unity for vanishing coupling  \cite{Kutasov:2004xu} and reads $F(\alpha)=1-2C^G_2\alpha$ in the NSVZ scheme  \cite{Novikov:1983uc,Novikov:1985rd} with $C^G_2$ the quadratic Casimir in the adjoint.  
Close to the Gaussian fixed point, the anomalous dimensions vanish and we can then read off the condition for a saddle
in terms of the  field multiplicities \eq{Ns}, 
\beq\label{Bs}
\begin{array}{lcl}
 3{N_1}&>&{N_2+N_F}\\[1ex]
3N_2 &\le &{N_1+N_F+N_Q}\,.
\end{array}
\eeq
This also covers the case where the one-loop coefficient of $\beta_2$ vanishes identically. For the Yukawa coupling, we exploit that  supersymmetry dictates strict non-renormalisation theorems which ensure that  superpotential couplings are only renormalised through field anomalous dimensions. Hence, the RG running  is given  by
\bea
\label{betay}
\beta_y 
&= &2\alpha_y\big[\gamma_\psi+ \gamma_{\Psi}+ \gamma_{\chi}\big]\,,
\eea
valid to all orders in perturbation theory. Renormalisation group fixed points $(\beta_i=0)$ correspond to superconformal field theories.

All perturbative  fixed points    $(\alpha^*_i\ll 1)$ for this theory have  previously been found in 
\cite{Bond:2017suy}. In general, these are either Banks-Zaks fixed points where some of the gauge couplings are non-zero but $\alpha^*_y=0$, or gauge-Yukawa (GY) fixed points with one $({\rm GY}_1)$, the other $({\rm GY}_2)$, or both gauge couplings non-zero $({\rm GY}_{12})$ and  $\alpha^*_y>0$ \cite{Bond:2016dvk,Bond:2018oco}. It has also been shown that if ${\rm GY}_1$ is UV, its outgoing trajectory is connected with the IR fixed point $({\rm GY}_{12})$, and similarly for  ${\rm GY}_2$  \cite{Bond:2017suy}.

In this work, we focus on the  regime \eq{Bs} and search for  non-perturbative fixed points  with the property
\beq\label{GY1}
{\rm GY}_1:\quad\quad  \alpha_{1}^*>0\,,\quad \alpha_{y}^*>0\,,\quad \alpha_2^*=0\,.
\eeq
Our {\it ansatz} states that the $SU(N_2)$ gauge sector is infrared free, meaning $\beta_2>0$ in the vicinity of the Gaussian. If $\beta_2<0$ in the vicinity of the
interacting fixed point,  the $SU(N_2)$ gauge sector  suddenly becomes asymptotically free, and the  fixed point is ultraviolet. 
In its vicinity, $\alpha_2$ is the only relevant coupling in the UV, which runs out of the fixed point as
\beq\label{relevant}
\alpha_2(\mu)=\frac{\delta\alpha_2(\Lambda)}{1+B_{2,\rm eff}\,\delta\alpha_2\,\ln(\mu/\Lambda)}\,,
\eeq
with $\delta\alpha_2(\Lambda)$ a small deviation at the high scale $\Lambda$, $B_{2,\rm eff}=B_{2} +N_F\,\gamma_\chi +4 N_1\,\gamma_\Psi>0$ the interaction-induced one loop coefficient, and $B_{2}<0$ given by (minus) the one-loop coefficient of \eq{beta2} at the Gaussian. The couplings $\alpha_1$ and $\alpha_y$ are irrelevant interactions in the UV and their running is fully determined by the one of $\alpha_2$ \cite{Bond:2017suy}. This  scenario is schematically depicted in Fig.~\ref{pSchematic}.

The  necessary and sufficient conditions for the partially interacting fixed point to turn the dangerously irrelevant coupling $\alpha_2$ into a relevant one are given by 
\bea
0&=&3N_1-N_F\left(1-2\gamma_\psi\right)-N_2\left(1-2\gamma_\Psi\right)
 \nonumber
 \label{NSVZ1}
\\
0&<&3N_2-N_F\,\left(1-2\gamma_\chi\right)
-N_1\left(1-2\gamma_\Psi\right)
-N_Q\left(1-2\gamma_Q\right)\,,
\label{NSVZ2}\\
0&= &\gamma_\psi+ \gamma_{\Psi}+ \gamma_{\chi}\,,
\label{NSVZy}
\nonumber
\eea
where the   two equations determine the  fixed point while the inequality ensures the sign flip from $\beta_2\ge 0$ close to the Gaussian to $\beta_2<0$ close to the   fixed point \eq{GY1}.

\section{a-Maximisation}

 The beta functions \eq{beta1}, \eq{beta2} and \eq{betay}, and the conditions \eq{NSVZ2}, still depend on the  superfield anomalous dimensions. These can  be determined either perturbatively, as we will do in Sec.~\ref{PT} below, or exactly, as we do here. To that end, we exploit that fixed points of the renormalisation group correspond to superconformal field theories. Hence superfields must transform in representations of the superconformal algebra. Focussing  on the bosonic part, we observe an extra global $U(1)_R$ symmetry in addition to the ordinary conformal algebra. The global and anomaly-free  $U(1)_R$ symmetry 
 prescribes global charges $R_i$ for all chiral superfields at any superconformal fixed point and  thereby determines anomalous dimensions $\gamma_i$  of chiral superfields via 
\beq\label{Ri}
\gamma_i=\0{3}{2}R_i-1\,.
\eeq
The task then reduces to the determination of $R$-charges at superconformal fixed points for which we use the method
of  $a$-maximisation \cite{Intriligator:2003jj}.  
Specifically, at  the fixed point GY${}_1$ with \eq{GY1} and characterised by \eq{NSVZ2},  we find
\beq\label{Rcharges}
\begin{array}{rl}
	R_\psi &\displaystyle=\frac23\left(1+\gamma_\psi\right)= 
	\frac{N_F[N_F - (N_1 + N_2)]^2 - N_1^2\, N_2\,\Delta}{[N_2 - N_F][N_1(N_2 + N_F) 
	+ (N_2-N_F)^2
     ]}\,,
	
	\\[2.5ex]
	R_\Psi &  \displaystyle
	=\frac23\left(1+\gamma_\Psi\right)=
	 \frac{N_2[N_2 - (N_1 + N_F)]^2 - N^2_1\,N_F\,  \Delta}{[N_F - N_2][N_1(N_2 + N_F) 
	 	+ (N_2-N_F)^2
      ]}\,,
\\ [2.5ex]
	R_\chi &=\displaystyle
\frac23\left(1+\gamma_\chi\right)=
	 	 \frac{8}{9}\frac{(N_2 - N_F)^2}{ (N_2 - N_F)^2-(1-\Delta) N_1^2}\,,
\\ [2.5ex]
	R_Q &\displaystyle
	=\frac23\left(1+\gamma_Q\right)=\frac23\,,
\end{array}
\eeq
and $\Delta > 0$  the positive root of
\bea	
\Delta^2 &=& 1+ \left(\frac{N_2-N_F}{3N_1^2}\right)^2\left[ (4N_1+N_2+N_F)^2 - 34N_1^2-4N_2N_F\right]\,.
\eea
The $\psi_Q$ fermions do not interact at the fixed point \eq{GY1}, hence $\gamma_Q=0$ and none of the $R$-charges depends on $N_Q$. The results for the $R$-charges uniquely determine the remaining anomalous dimensions non-perturbatively, and provide    closure for the beta functions  \eq{beta1}, \eq{beta2}  and \eq{betay}.  At weak coupling,  anomalous dimensions are small and $R$-charges are close to their classical values $(R_i\approx \s023)$. 
Moreover, unitarity mandates that scaling dimensions $D$ of spinless operators must satisfy $D\ge 1$  \cite{Mack:1975je}, additionally implying  $\gamma_i\ge -\s012$ ($R_i\ge \s013$)  for  gauge-invariant scalar operators  such as $\bar \psi \psi$. 

The conditions  for unitary quantum field theories  with interacting UV fixed points  \eq{NSVZ2} only depend on ratios of field multiplicities. 
Therefore, we can reduce the four-dimensional parameter space of field multiplicities \eq{Ns} to a three-dimensional one. Following  \cite{Bond:2017suy}, we do so by scaling-out  one of the field multiplicities, say $N_1$, and by introducing three suitable ratios  of field multiplicities instead, 
\beq\label{Peps}
\begin{array}{rcl}
R&=&
\displaystyle
\0{N_2}{N_1}\,,\quad
P=\0{N_1}{N_2}\0{N_Q+N_1+N_F-3N_2}{N_F+N_2-3N_1}\,,\quad 
\eps=
\displaystyle
\0{N_F+N_2}{N_1}-3\,.
\end{array}
\eeq
The colour ratio  $R$ (not to be confused with the $R$-charges $R_i$) is sensitive to the relative size of gauge groups. 
The parameter  $P<0$ is proportional to the ratio of one-loop gauge coefficients.
The parameter $\eps$, which in the region of interest is negative $\eps<0$, can be made arbitrarily small in a Veneziano limit where it controls  perturbative fixed points provided $|\eps|\ll 1$.
Since field multiplicities are semi-positive numbers, we further observe 
\beq\label{phys}
0<R<3\,,\quad P <\frac{4(1-R)+\eps}{R\,\eps}\,,\quad R-3<\eps<0\,.
\eeq
The parameters \eq{Peps} take discrete values for integer field multiplicities except  in
a Veneziano large-$N$ limit where they become continuous.

\section{Fixed Points and Conformal Windows} 

\begin{figure}
\begin{center}
\includegraphics[width=.9\textwidth]{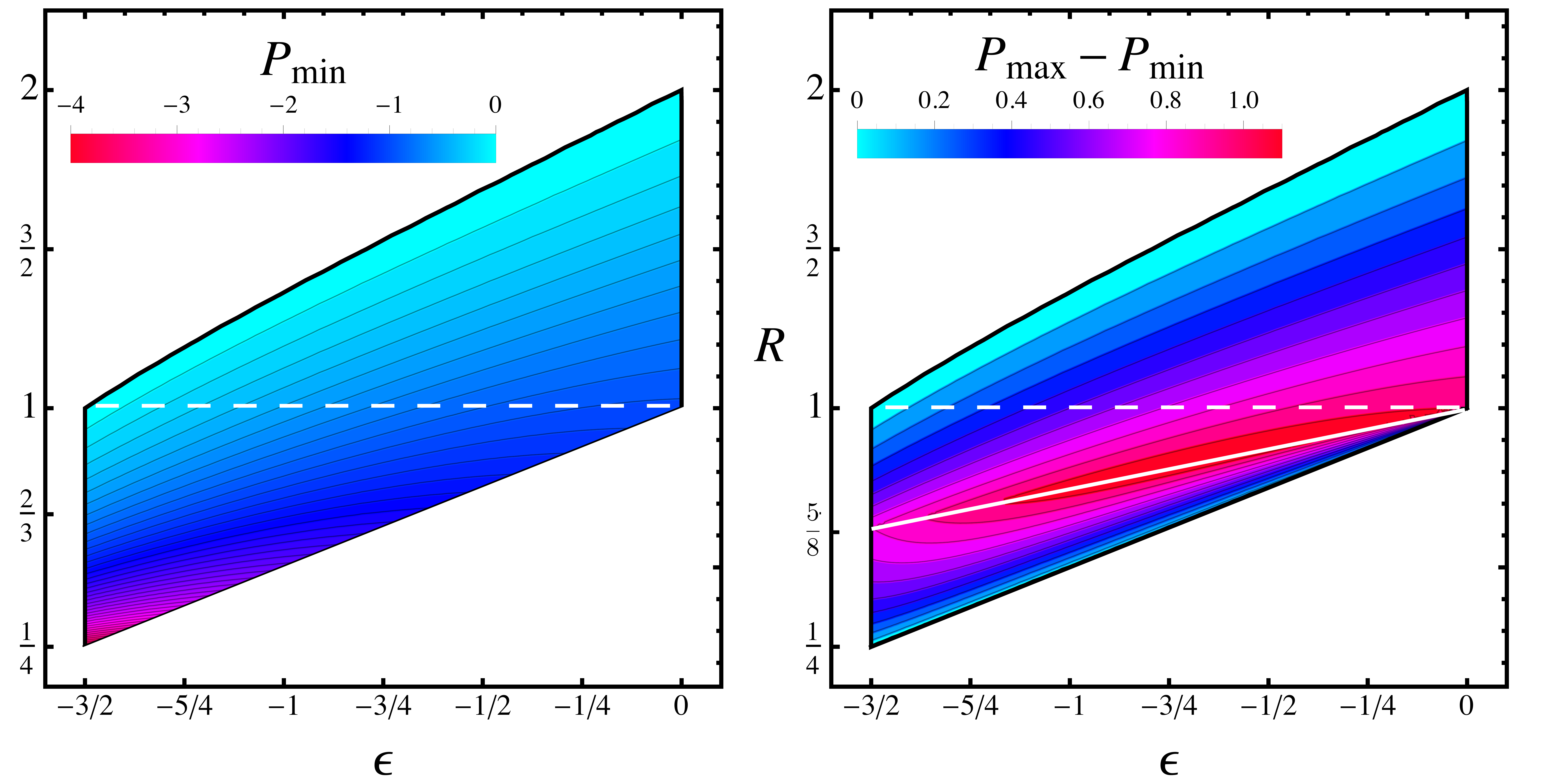}
\vskip-.2cm
\caption{Contour plot of  the superconformal window for asymptotic safety \eq{Prange} in the $(R,\eps)$ plane   showing the boundary $P_{\rm min}$ (left) and  the height $P_{\rm max}-P_{\rm min}$ (right); dashed and full white lines are explained in the main text.
 }\label{pConformalWindow}
\end{center}
\vskip-.6cm 
\end{figure}

We are now in a position to investigate the range of field multiplicities for which the theory displays an interacting ultraviolet fixed point.
Exploiting all constraints  
we find that the parameter $P$ is bounded from above and from below
\beq\label{Prange}
P_{\rm min}(R,\eps)<P\le P_{\rm max}(R,\eps)\,.
\eeq
The upper boundary $P_{\rm max}$ relates to \eq{Bs} and $P\le 0$, and to the physicality of 
field multiplicities \eq{phys}, whichever is stronger. The lower boundary  $P_{\rm min}$ relates to the sign-flip 
for induced asymptotic freedom  of $SU(N_2)$.  Explicitly,
\bea \label{Pmax}
P_{\rm max}(R,\eps)&=&
\left\{
\begin{array}{cl}
\displaystyle
\frac{4(1-R)+\eps}{R\,\eps}&\ \  {\rm for}\ \ R<1+\0{\eps}{4}\\[2ex]
0& \ \ {\rm otherwise}
\end{array}
\right.\\[1ex]
\label{Pmin}
P_{\rm min}(R,\eps)&=& \frac{3R_\Psi-2}{R\,\eps} 
+\frac{(3\,R_\chi-2)\,(3-R+\eps)}{R\,\eps}\,.
\eea
In the above,  the $R$-charges $R_\Psi$ and $R_\chi$ are   understood as functions of $(R,\eps)$ via  \eq{Rcharges} and  \eq{Peps}. 
The parameters $R$  and $\eps$ are  constrained  globally by  the physicality of couplings
 ($\alpha\ge 0$) and unitarity,
\beq \label{window-eps}
\begin{array}{rcl}
\s014<&R&<2\quad{\rm and}\quad
-\s032\le \eps<0\,.
\end{array}
\eeq
Fig.~\ref{pConformalWindow} shows a contour plot of the 
superconformal window \eq{Prange} and \eq{window-eps} in the $(\eps, R)$ plane.
The left panel shows the lower boundary $P_{\rm min}$ while the right panel shows the accessible range of $P$ values between $P_{\rm min}$ and $P_{\rm max}$. The lower boundary in both graphs is given by $R_{\rm min}=1+\s0{\eps}{2}$, corresponding to $N_1+N_2=N_F$. 
Above the full white  line $R=1+\s0{\eps}{4}$ we have $P_{\rm max}=0$. Roughly speaking, the width in $P$ is largest around the full white line. At the upper border we have $R_{\rm max}=2+\s023\eps$ which translates into $N_2=N_F$. We notice that the width $P_{\rm max}-P_{\rm min}$ vanishes at $R_{\rm min}$, while it remains small but non-zero at $R_{\rm max}$, except at the endpoints.
Weakly coupled  fixed points correspond to parameters $(P,R,\eps)$  close to the boundary  where $|\eps|\ll1$. 

\begin{figure}
\begin{center}
\vskip-0.2cm
\includegraphics[width=0.45\textwidth]{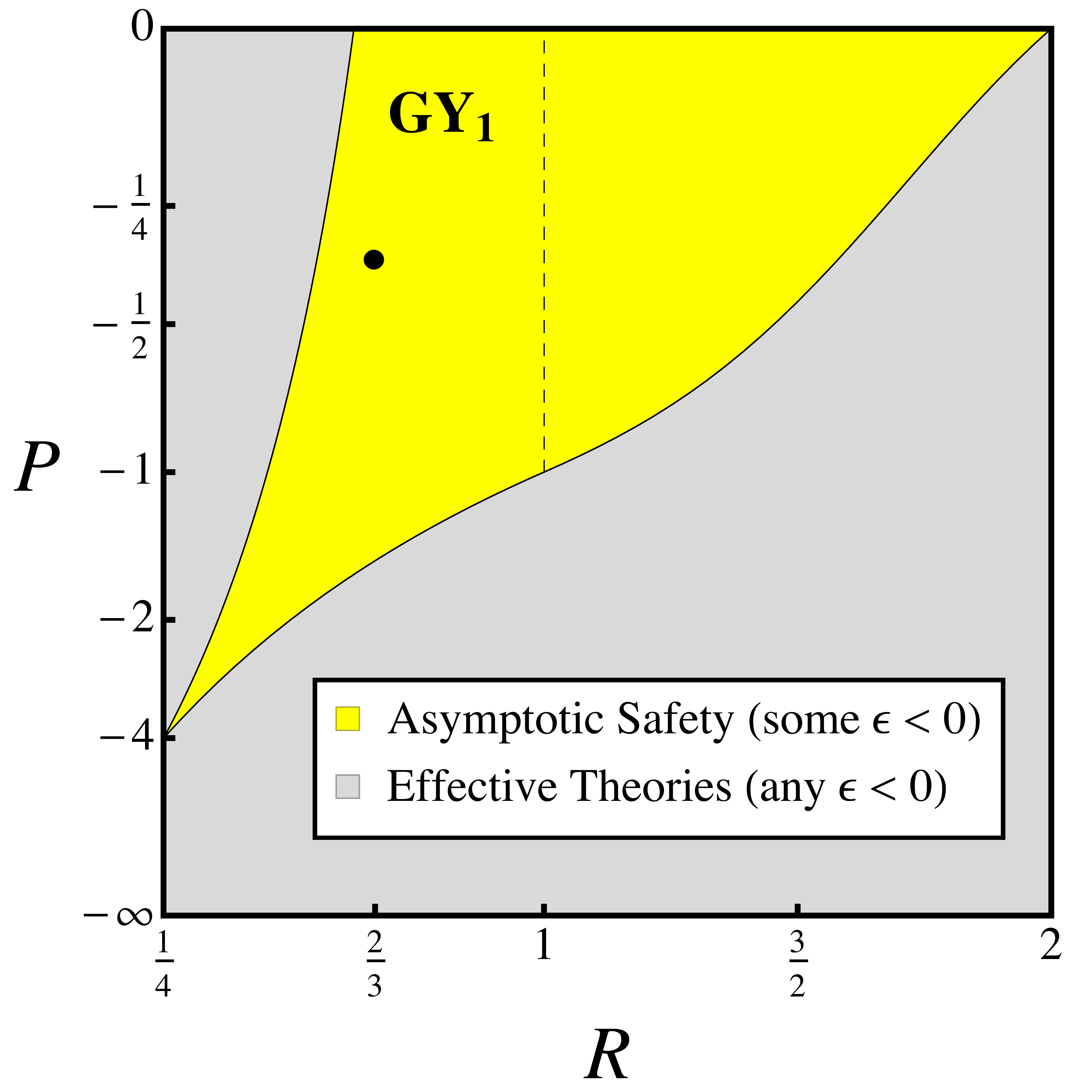}
\vskip-0.3cm
\caption{
Projection of the superconformal window with interacting ultraviolet fixed points    onto the $(R,P)$ parameter plane.  Weakly coupled fixed points arise within the yellow-shaded area on the right of the dashed line.  Strongly-coupled fixed points arise within the entire yellow-shaded area for some range of $\eps$. The full dot  is discussed in the main text.
\label{pAS_strong}
}
\end{center}
\vskip-.6cm 
\end{figure}

The conformal window is further illustrated in Fig.~\ref{pAS_strong} showing its projection onto the $(P,R)$ plane. Within the yellow-shaded area, weakly coupled fixed points arise  towards the right of the dashed line, while strongly-coupled fixed points can arise anywhere. Outside the yellow-shaded area, asymptotic safety is not available  and the corresponding quantum field theories must viewed as effective rather than fundamental. 

It is interesting to discuss the  $N_F$ dependence of fixed points within the conformal window
while keeping $(N_1, N_2,P)$ fixed.  In    Fig.~\ref{pConformalWindow}, this  effectively corresponds to varying $\eps$  along horizontal cuts. 
We notice that the conformal window splits into two distinct types of models, separated by  dashed  lines in Figs.~\ref{pConformalWindow} and \ref{pAS_strong},  to which we refer as ``mostly weakly" and ``mostly strongly" coupled.  Specifically, the first subset of models are those above the dashed line in Fig.~\ref{pConformalWindow} and on the right of the  dashed  line in Fig.~\ref{pAS_strong}. Their conformal window 
is characterised by
   \beq\label{CW1}
\begin{array}{rcl}
1\le &R&<2\,.
 \end{array}
  \eeq
    Then, for any admissible value of $R$ within \eq{CW1}, there is a range of viable parameters $P$ within $(-1,0]$ such that the UV conformal window in $\eps$ covers the range 
    \beq \label{CW1e}
    \eps_{\rm max}\le \eps<0\,.
    \eeq
    The lower bound $\eps_{\rm max}=\s032(R-2)$ may become as low as $-\s032$. The significance of this result is as follows. Perturbative fixed points are controlled by the parameter $|\eps|\ll 1$ in a large-$N$ Veneziano limit, meaning that this part of the UV conformal window
        contains all perturbatively controlled superconformal UV fixed points  found previously in \cite{Bond:2017suy}. 
    Hence,  we observe     that all perturbative fixed points extend into  a non-perturbative conformal window for $\eps$ given precisely by the range \eq{CW1e}.
   The same considerations  apply  for  finite $N$, away from a Veneziano limit, the only difference being that  $(R,P,\eps)$ take discrete rather than continuous values. Since all theses fixed points are linked to weakly coupled ones, and for want of terminology, we   refer to the models  within  \eq{CW1}    as  ``mostly weakly" coupled.

 Next, we turn to the part of the conformal window 
 below the dashed line in Fig.~\ref{pConformalWindow} and to the left of the  dashed  line in Fig.~\ref{pAS_strong}, characterised by
  \beq\label{CW2}
 \begin{array}{rcl}
 \s014<&R&<1\,.
\end{array} \eeq
For any admissible value of $R$ within \eq{CW2}, there is a range of viable parameters $P$ within $(-4,0]$ such that the UV conformal window in $\eps$ covers the range 
    \beq \label{CW2e}
   -\s032\le \eps<\eps_{\rm min}<0\,.
    \eeq
    Here, the   lower bound $\eps_{\rm min}=2(R-1)<0$ ensures induced asymptotic freedom for the coupling $\alpha_2$.
 Most importantly, we observe the existence of a finite gap $[\eps_{\rm min},0]$ in $\eps$  within which no  asymptotically safe fixed points can be found.  This implies that none of the fixed points within \eq{CW2}  can be achieved with strict perturbative control $(\eps\to 0^-)$, not even in a Veneziano limit.
 It is in this sense that the  fixed points within \eq{CW2}  are non-perturbative and parametrically disconnected  from the free theory, quite different from  those in \eq{CW1}.  For these reasons, we refer to these superconformal fixed points  as ``mostly strongly" coupled.

Another important aspect of the conformal window relates to the  saddle close to the Gaussian which needs to be overcome by the interacting fixed point.
Recall that the asymptotically free (infrared free) direction is characterised by the one-loop gauge coefficient $B_1=2 \,N_1\,\eps<0$  ($B_2=2\, N_2\,P\, \eps\ge 0$), with (minus)  their ratio  given by the imbalance parameter 
\beq\label{I}
I=-B_2/B_1\equiv |P|\cdot R
\eeq
for which  we have $I\ge 0$. Quantum effects at the interacting fixed point overcome the positive or vanishing one-loop coefficient $B_2$ and turn it, effectively, into a negative one (see Fig.~\ref{pSchematic}). It is then important to understand the largest  imbalance
that can be achieved  without spoiling asymptotic safety. We find that the imbalance is bounded from above,
\beq\label{PR}
 0\le I=\0{3N_2-N_1-N_F-N_Q}{N_F+N_2-3N_1}<1\,,
 \eeq 
where we recall that field multiplicities obey \eq{Bs}. In other words, as soon as one gauge sector is as or more infrared free at the Gaussian fixed point than the other gauge sector is ultraviolet free, asymptotic safety at an interacting fixed point cannot arise.

 In Figs.~\ref{pConformalWindow} and~\ref{pAS_strong}, the small imbalance region $0\le I\ll 1$ is realised close to the upper boundaries where $P\approx 0$.
Here, many ultraviolet fixed points can be found  in large parts of the parameter space  including perturbative and non-perturbative ones. 
    With growing imbalance $I\to 1$, the set of perturbatively controlled fixed points shrinks to the vicinity of a single point $(P,R)=(-1,1)$.\footnote{In a Veneziano limit, this are the models with $N_1,N_2,N_F\to\infty$ while $N_F/N_1=N_F/N_2=2$, and $N_Q/N_1\to 0$.} 
    Non-perturbatively, however, many more fixed points  realise a near-maximal imbalance, corresponding to  the lower boundary  in Fig.~\ref{pConformalWindow} with parameters  given by the  line  $R=R_{\rm min}(\eps)$,  $P=-1/R_{\rm min}$,  and  $\eps \in [-\s032,0)$.   In Fig.~\ref{pAS_strong}, the near-maximal imbalance is realised along the lower boundary to the left of the dashed line.  
 We conclude that all models which may afford the near-maximal imbalance are contained  in the ``mostly strongly coupled" part of the conformal window \eq{CW2}.

\begin{figure}[t]
\begin{center}
\vskip-0.2cm
\includegraphics[width=0.45\textwidth]{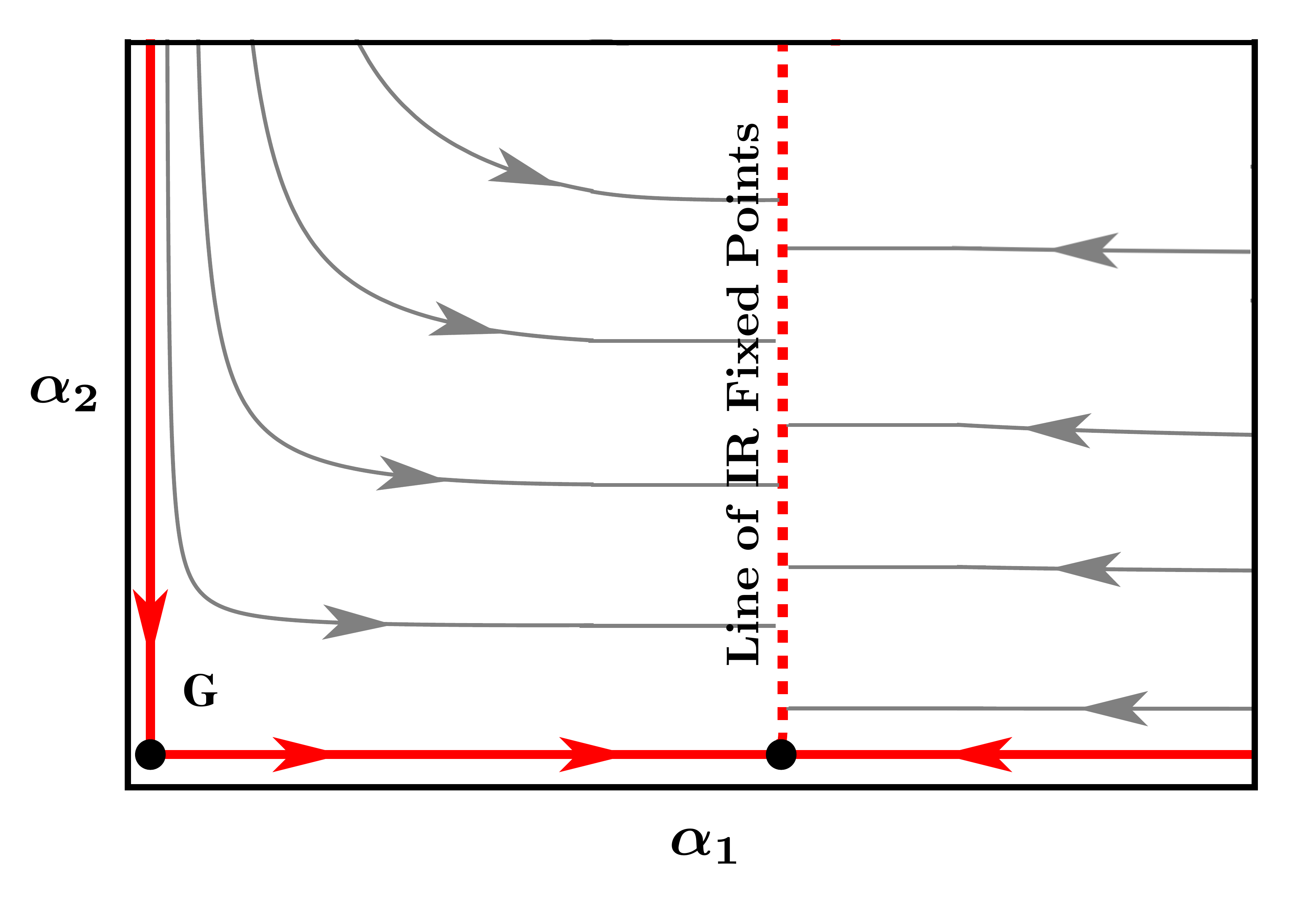}
\vskip-0.3cm
\caption{
Schematic phase diagram of Leigh-Strassler type models in the plane of gauge couplings in the limit where the UV fixed point degenerates into a line of IR fixed points (see Fig.~\ref{pSchematic}).\label{pSchematicLine}
}
\end{center}
\vskip-.6cm 
\end{figure}

We emphasize  that the boundary at $P_{\rm min}$ in \eq{Pmin} 
is not part of the conformal window. The reason for this is that whenever $P\to P_{\rm min}$
the inequality in \eq{NSVZ2} becomes an equality, meaning that $\beta_2\to 0$ non-perturbatively at the fixed point \eq{GY1}. At this point the UV fixed point  ${\rm GY}_1$ merges with yet another fixed point (a non-perturbative IR fixed point ${\rm GY}_{12}$  \cite{Bond:2017suy}) in the limit $P\to P_{\rm min}$. In consequence, the fixed point \eq{GY1}  degenerates into a line of   fixed points for any $\alpha_2$, illustrated in Fig.~\ref{pSchematicLine}. 
As such, this limit offers a manifold of Leigh-Strassler type models  \cite{Leigh:1995ep}, each of which is characterised by a 
line 
of interacting superconformal field theories, disconnected from the free theory, and generated by an exactly marginal operator. Further, each of these lines of fixed points corresponds to an infrared sink because the previously relevant perturbation,  given by $\alpha_2$, has become strictly marginal. 
This  includes all settings with maximal imbalance $I=1$. Concrete examples for strongly-coupled Leigh-Strassler models 
are delegated to Sec.~\ref{benchmarks} below.\footnote{The fate of an IR sink can be evaded by adding mass term perturbations for the $N_F$ superfields $\psi$ which may lift the degeneracy and allow RG trajectories to emanate from the line of fixed points. However, this  mechanism   defies the original setup \eq{Bs} in that the removal of the $\psi$ degrees of freedom would make the theory asymptotically free from the outset.}

\begin{figure*}
\begin{center}
\includegraphics[width=.9\textwidth]{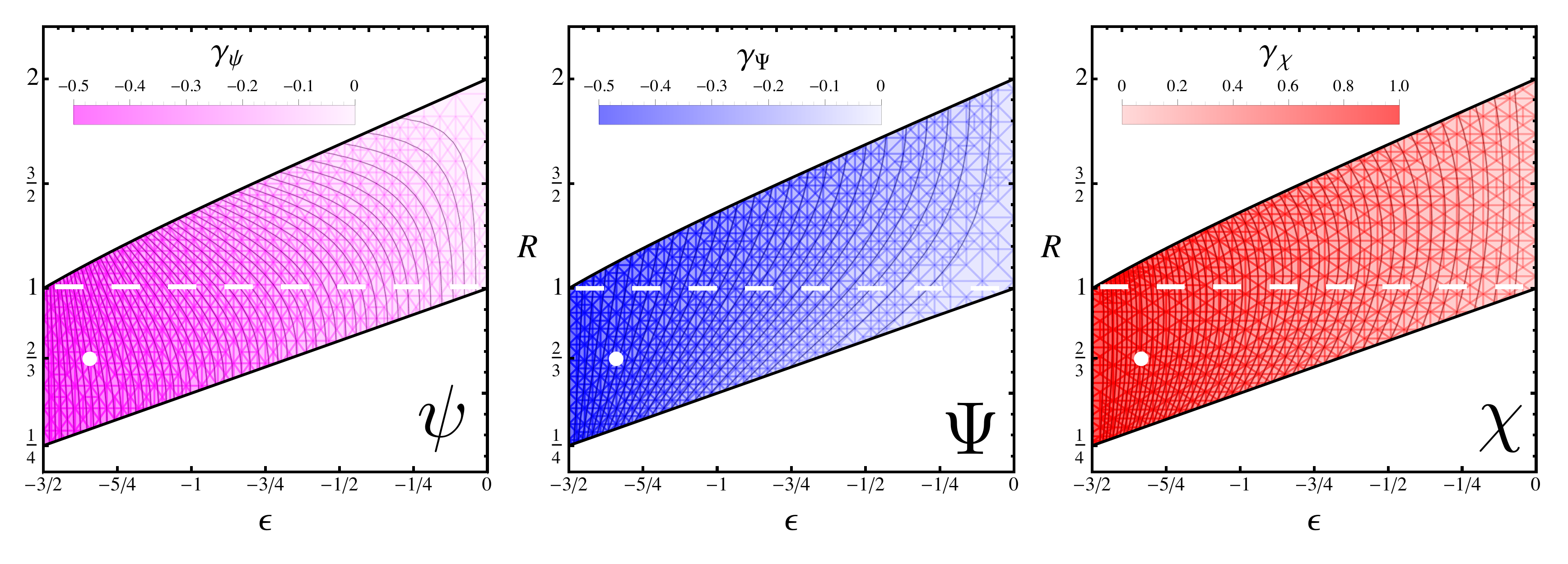}
\vskip-.5cm
\caption{
Contour plot showing the non-perturbative anomalous dimensions of chiral superfields  
at  the superconformal fixed points 
of Fig.~\ref{pConformalWindow}.  The full dot and dashed line are explained in the main text.\label{pGamma}
}
\end{center}
\vskip-.6cm 
\end{figure*}

 Fig.~\ref{pGamma} shows the superfield anomalous dimensions within the entire conformal window. Since the spectator fermions $\psi_Q$ are free at  the fixed point \eq{GY1}, the chiral superfield  anomalous dimensions   are only functions of $(R,\eps)$ and  independent of $P$.        
 Overall, we find that anomalous dimensions grow with growing $|\eps|$. For models within \eq{CW1} or \eq{CW2}, the chiral anomalous dimensions cover the  range 
 \beq\label{grange}
 0>\gamma_\psi,\gamma_\Psi\ge -\s012\quad{\rm and}\quad 0<\gamma_\chi\le 1
 \eeq
 with extremals reached at the $\eps=-\s032$ and $\eps=0$ boundaries of the conformal window. For models within \eq{CW2} anomalous dimensions tend to take larger values than for those in  \eq{CW1}. A comparison with  perturbation theory is given in Sec.~\ref{PT}. Incidentally, the monotonicity of $\gamma_i$ with growing $\eps$ establishes that anomalous dimensions do not vanish unless the fixed point is free. 
 
\section{Central Charges and the $a$-Theorem} 
Superconformal fixed points can also be characterised by  central charges $a$, $b$ and $c$,  the anomaly coefficients  \cite{Anselmi:1997am,Anselmi:1997ys}. Their 
values have to satisfy certain conditions, and can be used to constrain viable fixed points.
The global charges $a$ and $c$ can be expressed in terms of the $R$-charges of chiral superfields,
 \begin{align}\label{eq:afunc}
	a &= \frac{3}{32}\left(3\Tr R^3 - \Tr R\right)\,\\
	c &= \frac{1}{32}\left(9\Tr R^3 - 5\Tr R\right)\,.
\end{align}
Given that the  positivity conditions $a,b,c>0$ are satisfied non-marginally for free theories,  they are guaranteed to be  satisfied for perturbative theories. 
Similarly, the conformal collider bound \cite{Hofman:2008ar}
\beq
\012<\frac{a}{c}<\032
\eeq
cannot be violated for weakly interacting theories. 
Here, we find that the  positivity conditions and the conformal collider bound are  satisfied  non-perturbatively,
in the entire conformal window, as they must. 

We now turn to the $a$-theorem. It states that the central charge $a$ must be a decreasing function along RG trajectories in any 4d quantum field theory  \cite{Komargodski:2011vj,Komargodski:2011xv}. We find that
$a_{\rm UV}-a_{\rm Gauss} <0\,,$
which  confirms that none of the UV fixed points is connected by an RG trajectory with the free Gaussian fixed point, in accord with Fig.~\ref{pSchematic}. Further, we noted earlier that any interacting UV fixed point in the conformal window comes bundled with a  
fully interacting non-perturbative IR fixed point $({\rm GY}_{12})$ \cite{Bond:2017suy}. We find
$a_{\rm UV}-a_{\rm IR} >0\,,$ meaning that  both conformal fixed points are connected by RG trajectories  flowing from  the former to the latter as in Fig.~\ref{pSchematic}.  We conclude that our results are consistent with the  $a$-theorem, as they must.

\section{Seiberg Duality}

 Next, we   comment on how our results relate to Seiberg's electric-magnetic duality \cite{Seiberg:1994pq}. 
At the partially interacting fixed point \eq{GY1} where the $SU(N_2)$ gauge sector is free, the flavour symmetry  of the theory is enhanced. This manifests itself as an exchange symmetry under $N_2 \leftrightarrow N_F$, whereby the quarks $\psi$ and $\Psi$ interchange their roles while the fields $\chi$ remain unchanged,
\beq
	N_2 \leftrightarrow N_F\,, \quad\quad R_\psi \leftrightarrow R_\Psi\,,	\quad\quad	R_\chi \leftrightarrow R_\chi\,,
\eeq
see \eq{Rcharges}.  Hence, we observe an interacting ``magnetic'' $SU(N_1)$ gauge theory which has $N_F+N_2$ flavours of ``magnetic quarks'' $\psi$ and $\Psi$, alongside $(N_F+N_2)^2$ singlet mesons $(\psi_L\psi_R)$, $\chi_{L}$, $\chi_{R}$, and $(\Psi_L\Psi_R)$ \cite{Barnes:2005zn}. In this light, the gauge-Yukawa fixed point \eq{GY1} corresponds to a free non-Abelian magnetic phase whose superconformal window is known to cover the range 
$\s032 N_1<N_F+N_2<3N_1$ \cite{Seiberg:1994pq}
which corresponds to
 the parameter range 
$-\s032<\eps<0$ observed in \eq{window-eps}. Notice however that the conformal window in Fig.~\ref{pConformalWindow} has additional constraints  on $(R,P)$ which ensure that the $SU(N_2)$ gauge sector becomes a relevant perturbation at the fixed point. 

Seiberg duality predicts the existence of a dual ``electric'' theory which must have gauge group $SU(N_F+N_2-N_1)$ coupled to $N_F+N_2$ flavours of electric quarks. The conformal window of the non-Abelian Coulomb phase is given by 
 $\s032 (N_F+N_2-N_1)<N_F+N_2<3(N_F+N_2-N_1)\,.$  This corresponds to the exact same parameter band in $\eps$ as the one in \eq{window-eps}. 
In the special case where $N_F=N_2$, given by  the upper boundary in Fig.~\ref{pConformalWindow}, the $R$-charges simplify and become
\beq
R_\psi=R_\Psi=1-\frac{N_1}{2N_F}\,,\quad R_\chi=\frac{N_1}{N_F}\,.
\eeq
One recognises the $R$-charges for the quarks and singlet mesons of supersymmetric magnetic QCD  \cite{Seiberg:1994pq}, whose Seiberg dual is characterised by electric quarks ${\widetilde\psi}$ and $\widetilde\Psi$ with
\beq
R_{\widetilde\psi}=R_{\widetilde\Psi}=\frac{N_1}{2N_F}\,.
\eeq
We refer to  \cite{Barnes:2005zn} for further aspects of Seiberg duality in theories with product gauge groups.
 
\section{Perturbation Theory} \label{PT}

In this section, we contrast the non-perturbative determination of anomalous dimensions with perturbation theory, using general expressions for perturbative beta functions  in the DRED regularisation scheme  up to three loop  \cite{Einhorn:1982pp,Machacek:1983tz}. The main addition are perturbative expressions for anomalous dimensions which  at superconformal fixed points can be used to cross-check the non-perturbative results from $a$-maximisation.

For the sake of this comparison, it is convenient to perform a Veneziano limit and rescale gauge couplings as $\alpha_i\to N_i\,\alpha_i$ and Yukawas as $\alpha_y\to N_1\,\alpha_y$. We also use the parametrisation \eq{Peps}. 
The parameter $0<|\eps|\ll 1$ then serves as a small expansion parameter to ensure rigorous control of fixed points
in perturbation theory.  
To find   fixed points, scaling exponents, and anomalous dimensions to first (second) order in $\eps$, 
we must  retain  terms up to two (three) loop in the gauge coupling, and up to one (two) loop in the Yukawa beta functions \cite{Bond:2016dvk,Bond:2017tbw}.
We refer to these approximations as NLO and NNLO, respectively. Using the general results  of \cite{Einhorn:1982pp,Machacek:1983tz} 
we find the gauge beta functions up to three loop for our models as
\bea
	\beta_1^{(1)} &=& 
	\displaystyle
	2\alpha_1^2\, \eps 
\nonumber
\,,\\
	\beta_1 ^{(2)} &=& 
	\displaystyle
	2\alpha_1^2\big[6\alpha_1 + 2R \alpha_2 
	- 4R(3 +\eps- R)\alpha_y\big]
\label{beta13}
\,,\\ \nonumber
	\beta_1^{(3)} &=& 4\alpha_1^2\left[2\eps\alpha_1^2 
			 - R\left(2\alpha_1\, \gam{\Psi}{1}+\gam{\Psi}{2}\right)
			 -(3 +\eps- R)\left(2\alpha_1\,\gam{\psi}{1} +\gam{\psi}{2}\right)\right]\,,
			 \eea
and		 
\bea\nonumber
	\beta_2^{(1)}  &=& 
		\displaystyle
2\alpha_2^2\,P\eps
\,,\\
\label{beta23}
	\beta_2^{(2)}  &=& 
		\displaystyle
2\alpha_2^2\left[6\alpha_2 + \02R\alpha_1  
		- \04R(3 - R+\eps)\alpha_y\right]
\,,\\ \nonumber
	\beta_2^{(3)} &=& 
	4\alpha_2^2\left[2 P \eps\alpha_2^2 -\frac1R  \left(2\alpha_2\gam{\Psi}{1}+  \gam{\Psi}{2}\right)
\right.\\&&\nonumber 	\quad\quad \left. 
	- \frac{3-R+\eps}{R}\left( 2\alpha_2\,\gam{\chi}{1} +\gam{\chi}{2}\right)
- \left(4 + P \eps - \frac{4+\eps}{R}\right)\left( 2\alpha_2\,\gam{Q}{1}+\gam{Q}{2}\right)\right]\,.
			\eea
The perturbative Yukawa beta function is given by \eq{betay} for any loop order.
The anomalous dimensions of the superfields are required up to two loop accuracy. 
They read  
\beq\label{gamma1}
\begin{array}{rcl}
	\gam{\psi}{1} &= &\displaystyle
	R\,\alpha_y - \alpha_1\,,\\[1ex]
	\gam{\Psi}{1} &=&\displaystyle (3 - R+\eps)\alpha_y - \alpha_1 -\alpha_2\,,\\[1ex]
	\gam{\chi}{1} &=&\displaystyle \alpha_y - \alpha_2\,,\\[1ex]
	\gam{Q}{1} &= &\displaystyle-\alpha_2\,,
\end{array}
\eeq
and 
\beq\label{gamma2}
\begin{array}{rcl}
	\gam{\psi}{2} &= &\displaystyle-R\,\alpha_y\left(\gam{\Psi}{1} + \gam{\chi}{1}\right) - \alpha_1 \gam{\psi}{1} + 4\,\eps\,\alpha_1^2\,,\\[1ex]
	\gam{\Psi}{2} &=&\displaystyle -(3 - R+\eps)\,\alpha_y\left(\gam{\psi}{1} + \gam{\chi}{1}\right) - (\alpha_1 + \alpha_2)\gam{\Psi}{1}
				+ 4\,\eps\,\alpha_1^2 + 4 \,P\,\eps\,\alpha_2^2\,,\\[1ex]
	\gam{\chi}{2} &= &\displaystyle-\alpha_y\left(\gam{\psi}{1} + \gam{\Psi}{1}\right) - \alpha_2\gam{\chi}{1}
				+ 4\, P\,\eps\,\alpha_2^2\,,\\[1ex]
	\gam{Q}{2} &= &\displaystyle-\alpha_2\, \gam{Q}{1} + 4 \,P\,\eps \,\alpha_2^2\,,
\end{array}
\eeq
respectively. Then,  interacting fixed points with \eq{GY1} are determined via a Taylor expansion of couplings in $\eps$ around the zeros  of \eq{beta13} 
and \eq{betay},  also using \eq{gamma1} and \eq{gamma2}. Notice that \eq{beta23} is only used to determine whether
the sign change from $\beta_2>0$ at the Gaussian to $\beta_2<0$ at \eq{GY1} has taken place.
Inserting the perturbative results for the fixed point into  \eq{gamma1} and \eq{gamma2} provides explicit expressions
for the anomalous dimensions of the form
    \beq\label{gammaAs}
     \gamma_i(\eps)=A^{(1)}_{i}\,\eps+A^{(2)}_{i}\,\eps^2+{\cal O}(\eps^3)\,.
     \eeq
The six coefficients $A^{(n)}_{i}$  for $i=\psi, \Psi, \chi$ which for $n=1,2$  arise from the perturbative fixed point solutions at NLO and NNLO accuracy  \cite{Bond:2017lnq,Bond:2017tbw} still depend on the parameter $R$ (but not on $P$). Their explicit expressions are not given because they do not offer further insights for the present purposes.
  
  The expressions \eq{gammaAs}  must be compared with the exact $R$-charges from $a$-maximisation \eq{Rcharges}.  Writing the exact anomalous dimensions in terms of the  $R$-charges, and expanding  the expressions to second order in $\eps$ with the help of  \eq{Peps}, we find the identical result for the coefficients $A^{(n)}_{i}$ in \eq{gammaAs}. This establishes consistency of findings between $a$-maximisation and perturbation theory, as it must.

       \begin{figure}
\begin{center}
\includegraphics[scale=.4]{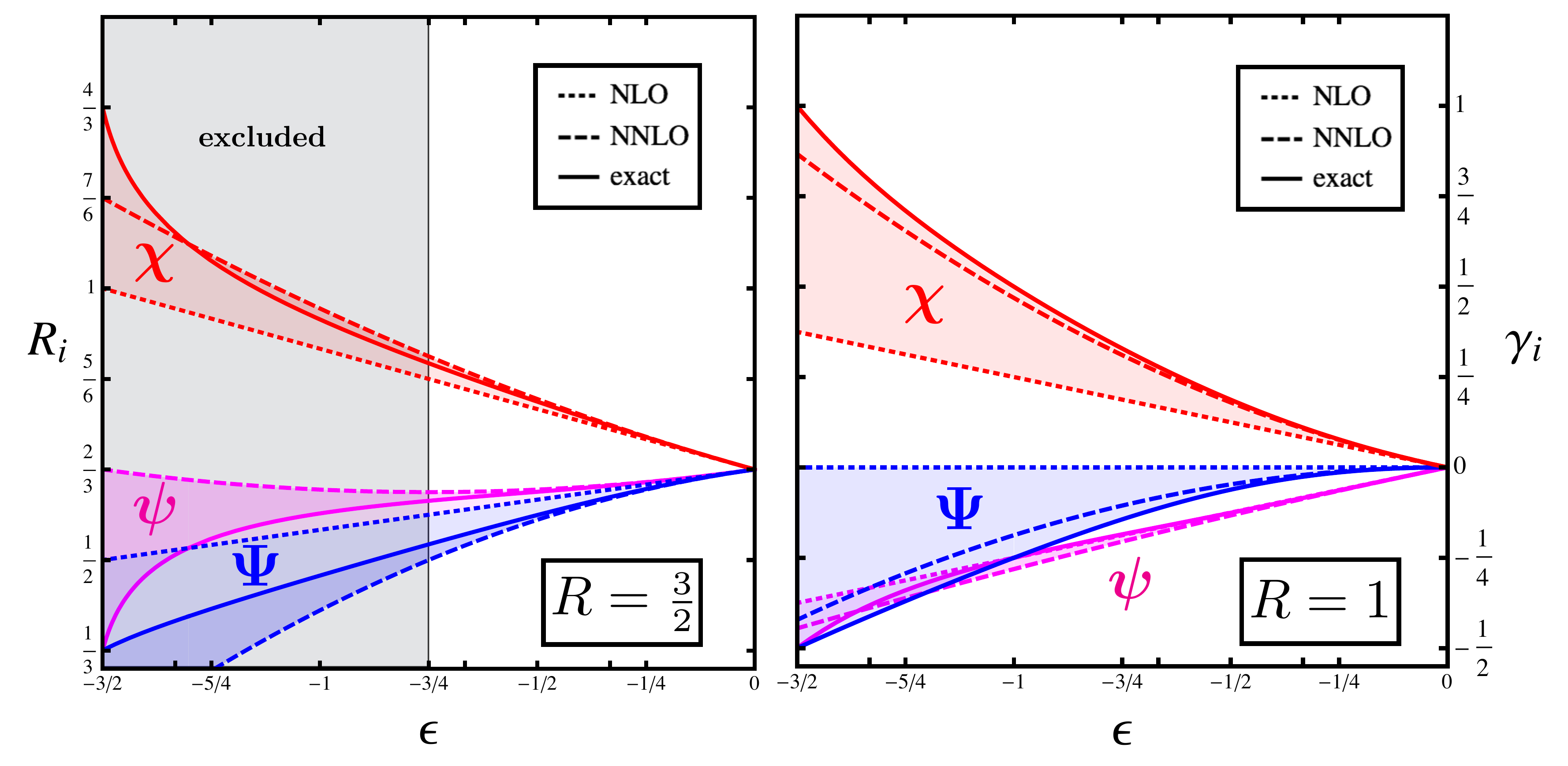}
\vskip-.3cm 
\caption{
Exact anomalous dimensions and $U(1)_R$ charges  in comparison with 2-loop (NLO) and 3-loop (NNLO) results from perturbation theory. Shown are projections along $R=\s032$ (left panel) and $R=1$ (right panel).  Fixed points in the gray-shaded area do not show the required sign-flip in \eq{NSVZ2} for any $P$, and are therefore outside of the UV conformal window.}
\label{pComparison1}
\end{center}
\vskip-.7cm 
\end{figure}

\section{Anomalous Dimensions}

\begin{figure}
\begin{center}
\includegraphics[scale=.4]{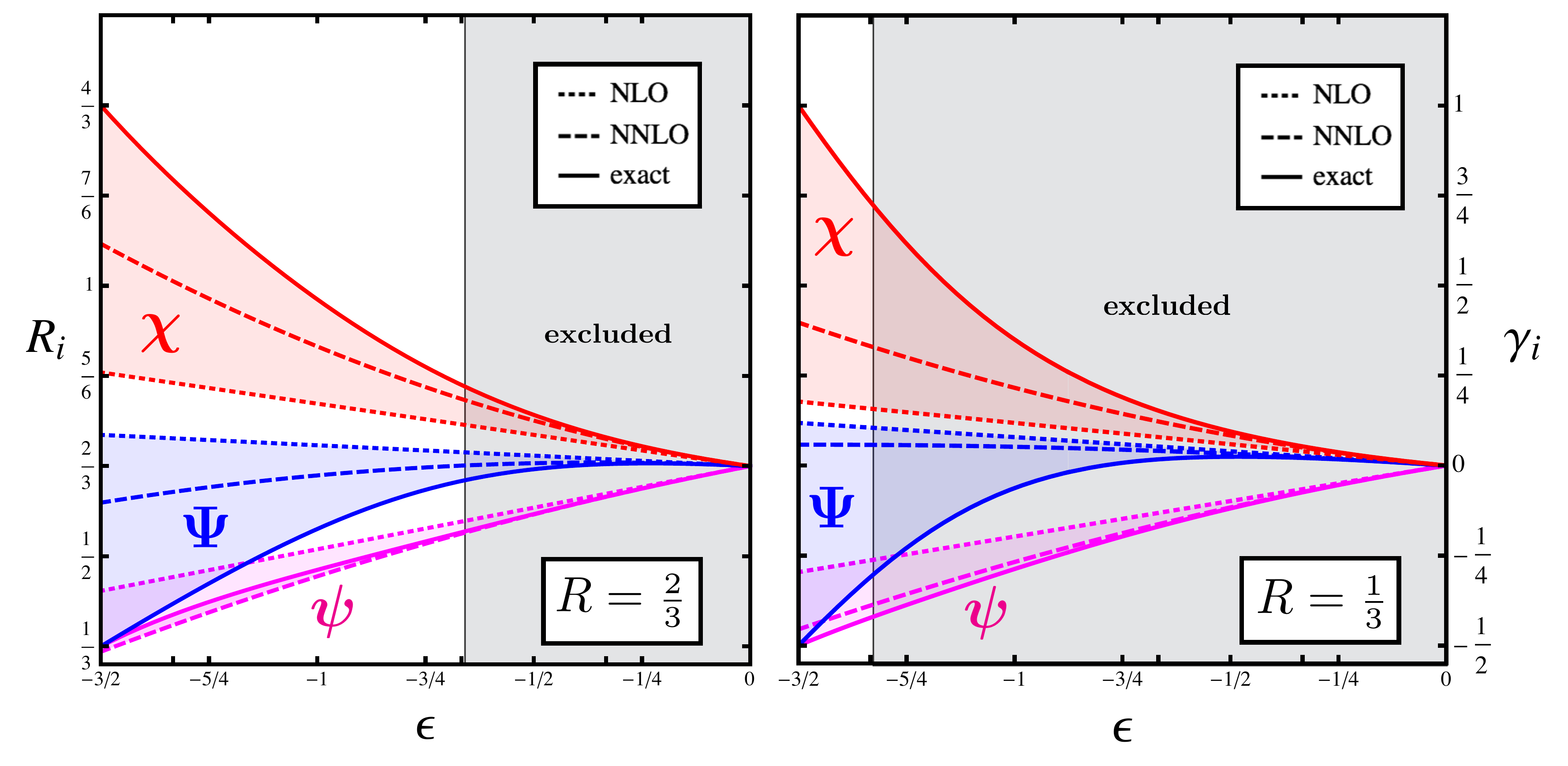}
\vskip-.3cm 
\caption{
Same as   Fig.~\ref{pComparison1}, except that the  projections are along $R=\s023$ (left panel) and $R=\s013$ (right panel). Decreasing $R$ considerably narrows the UV conformal window. The excluded (gray-shaded) areas refer to regions where some matter field field multiplicities would be unphysical (negative).}\label{pComparison2}
\end{center}
\vskip-.7cm 
\end{figure}

      In  Fig.~\ref{pGamma} we have shown the exact anomalous dimensions across  the entire UV conformal window. Here, we have a closer look into anomalous dimensions and compare with the  NLO and NNLO predictions from perturbation theory. More specifically, in Figs.~\ref{pComparison1}  and~\ref{pComparison2}, we  compare results 
      at fixed colour ratio $R=\s032, 1, \s023$ and  $\s013$, and  for any $\eps$ within $[-\s032,0]$, which corresponds to  horizontal cuts across Figs.~\ref{pConformalWindow} and~\ref{pGamma}.

       We begin with Fig.~\ref{pComparison1} where our results for $R=\s032$ (left panel) and $R=1$ (right panel) are shown. For small $|\eps|\ll 1$,  perturbation theory matches the exact results as it must. With growing $|\eps|$, all anomalous dimensions grow in magnitude. For $R=\s032$, the gray-shaded area indicates that fixed points are no longer in the UV conformal window, leading to a lower bound for $\eps$, see \eq{CW1e}. Also, the NLO results for $\gamma_\psi$ and $\gamma_\Psi$ coincide accidentally. At NNLO, perturbative results for all three anomalous dimensions are quite close to the exact results in the admissible range for $\eps$. For $R=1$,
      the extension of the conformal window is maximal. 
      Here, anomalous dimensions 
      correspond to 
        the fixed points along the dashed lines in Figs.~\ref{pConformalWindow},~\ref{pAS_strong} and \ref{pGamma}, which marks the boundary between the ``mostly weakly" and ``mostly strongly" coupled quantum field theories, see  \eq{CW1} vs \eq{CW2}. Maximal values $\gamma_{\psi,\Psi}\to -\s012$ and $\gamma_\chi\to 1$  are reached for $\eps\to -\s032$. 
              For $\gamma_\psi$, we observe that the NLO and NNLO results are close to the exact one over the entire range for $\eps$.
        For $\gamma_\Psi$, the NLO correction vanishes accidentally. Similarily, for $\gamma_\chi$, the NLO result strongly underestimates the exact value with growing $|\eps|$. However, at NNLO, perturbative results for all three anomalous dimensions are close to the exact findings in the entire range of $\eps$.  We conclude that differences between  NNLO      and exact results are  moderate over the entire interval. Note  that this does  not hold true in general, but when it does, one may use this near-coincidence to extract  estimates for  fixed points at strong coupling which elsewise are  not easily accessible.
        
        The same analysis is repeated in Fig.~\ref{pComparison2} for the parameter choices $R=\s023$ (left panel) and $R=\s013$ (right panel).  These horizontal cuts through Fig.~\ref{pConformalWindow} project onto the more strongly coupled fixed points in \eq{CW2}. For either of these  the small $\eps$ regions are excluded (grey-shaded areas) because the corresponding fixed points are not ultraviolet. In the increasingly narrow regions of interest \eq{CW2e} which are $\eps\in[-\s032,-\s023]$ and $\eps\in[-\s032,-\s043]$, respectively,  we  observe that differences between NLO, NNLO, and exact results become rather large for
        $\gamma_\chi$, $\gamma_\psi$ and $\gamma_\Psi$. Outside the UV conformal window, we also notice that $\gamma_\Psi$ is no longer monotonous with $\eps$,  but instead changes sign for non-vanishing $\eps$, and taking values outside the range \eq{grange}. Perhaps unsurprisingly, this confirms that the leading orders of perturbation theory cease to offer a good approximation for anomalous dimensions at strong coupling, once  $|\eps|$  is large.  

 \section{Benchmarks for Models of Particle Physics}\label{benchmarks}
 
At weak coupling, ultraviolet fixed points are often characterised by a large number of field multiplicities  \cite{Bond:2017suy}. At strong coupling, fixed points can arise with a much lower number matter fields. Therefore, we benchmark models with minimal numbers of gauge and matter fields  and  explore whether variants could serve as  templates  for asymptotically safe Standard Model extensions.

Our benchmarks of superconformal fixed points 
with low numbers of fields are summarised in Tab.~\ref{tsolutions}.
They represent strongly coupled models in that they have large  $|\eps|$ and 
anomalous dimensions. The benchmarks  also cover the full range of imbalance parameters $I$.
Note that because the $\psi_Q$ fermions are free at the superconformal fixed point, anomalous dimensions of chiral 
superfields agree between different models as long as they only differ in the $N_Q$ multiplicity, see models 2--4, model 5 and 6, 
or model 7 and 8.

We begin with model 1 which is the sole example with $SU(2)\times SU(2)$ gauge 
symmetry, the simple reason being that the only other integer solution $(N_F,N_Q)=(3,2)$
to the constraint  \eq{Bs} does not lead to asymptotic safety. Here, anomalous dimensions are still small and well-approximated by the perturbative three loop result  (see Fig.~\ref{pComparison1}, right panel). Moreover, in models 1 and 2, the $SU(2)$ gauge sector has a vanishing one-loop coefficient.
In both cases a viable interacting UV fixed point is found, located at a boundary of the conformal window (Fig.~\ref{pConformalWindow}) 
where $I=0$. 

For the gauge group $SU(3)\times SU(2)$, we find a total of five solutions (models 2--6).
For all of these  $SU(2)$ is infrared free at the Gaussian. Settings where $SU(3)$ is infrared free do not lead to asymptotic safety. 
Also, all models are within the strongly-coupled domain where 3-loop perturbation theory does not offer an accurate approximation  (Fig.~\ref{pComparison2}, left panel). Model 3 has $N_F=3$ and $N_Q=1$ and its location in the conformal window is  indicated by a full dot in Figs.~\ref{pConformalWindow} and~\ref{pAS_strong}. The anomalous dimensions are of order unity 
\beq
(\gamma_\psi,\gamma_\Psi,\gamma_\chi)\approx(-0.41, -0.38, 0.79)
\eeq 
and shown by a full dot in Fig.~\ref{pGamma}. Enhancing $N_Q$ by one unit  gives another viable fixed point  (model 4) without changing  the anomalous dimensions. For $(N_F,N_Q)=(4,0)$ (model 5) and $(N_F,N_Q)=(4,1)$ (model 6), the main change is that $|\eps|$, and hence the anomalous dimensions come out slightly smaller than in model 2-4. Overall, increasing the number of matter fields charged under $SU(2)$ also increases the imbalance parameter.  

Models $3-6$ have some similarities with the Minimal Supersymmetric Standard Model (MSSM). Unlike in the Standard Model, the  $SU(2)$ sector of the MSSM   becomes infrared free due to extra gauge charges from supersymmetry partners, while the $SU(3)$ sector  remains asymptotically free.
Hence, the Gaussian corresponds to a saddle with imbalance  $I_{\rm MSSM}=\s013$, just  as in model 5. The imbalance could even be made larger such as in model 4 or model 6, and still yield an asymptotically  safe fixed point. Hence, we see that supersymmetric gauge theories with the SM gauge groups (neglecting hypercharge) and  imbalances similar to the MSSM may very well become asymptotically safe
\cite{Hiller:2022hgt}.

\begin{table}[t]
 \aboverulesep = 0mm
\belowrulesep = 0mm
	\addtolength{\tabcolsep}{2pt}
	\setlength{\extrarowheight}{2pt}
\begin{center}
\begin{tabular}{`cc cc cc cc cc c`}
\toprule
\rowcolor{Yellow}
\bf \ Model\ \ 
&\bf \ Gauge Group \  
&\ \ $\bm{N_F}$\ \  
&\ \ $\bm{N_Q}$\ \  
&\ \ $\bm{I}$ \ \ 
&\ \ $\bm{R}$ \ \ 
&\ \ $\bm{P}$ \ \ 
&\ \ $\bm{\eps}$ \ \ 
&\ \ $\bm{\gamma_\psi}$ \ \
&\ \ $\bm{\gamma_\Psi}$ \ \
&\ \ $\bm{\gamma_\chi}$ \ \
\\
\rowcolor{Yellow}
&& && && && &&\\[-4mm]
\rowcolor{white}
\midrule
1
&$SU(2)\times SU(2)$
&3
&1
&$0$
&$1$
&$0$
&$-\s012$
&$-0.126$
&$-0.061$
&$0.187$\\
2
&$SU(3)\times SU(2)$
&3
&0
&$0$
&$\s023$
&$0$
&$-\s043$
&$-0.412$
&$-0.382$
&$0.794$
\\
3
&$SU(3)\times SU(2)$
&3
&1
&$\s014$
&$\s023$
&$-\s038$
&$-\s043$
&$-0.412$
&$-0.382$
&$0.794$\\
4
&$SU(3)\times SU(2)$
&3
&2
&$\s012$
&$\s023$
&$-\s034$
&$-\s043$
&$-0.412$
&$-0.382$
&$0.794$
\\
5
&$SU(3)\times SU(2)$
&4
&0
&$\s013$
&$\s023$
&$-\s012$
&$-1$
&$-0.288$
&$-0.174$
&$0.462$
\\
6
&$SU(3)\times SU(2)$
&4
&1
&$\s023$
&$\s023$
&$-1$
&$-1$
&$-0.288$
&$-0.174$
&$0.462$
\\
7
&$SU(6)\times SU(2)$
&7
&0
&$\s079$
&$\s013$
&$-\s073$
&$-\s032$
&$-\s012$
&$-\s012$
&$1$
\\&& && && && &&\\[-4.5mm]
\midrule
\rowcolor{LightGray}
$8$
&$SU(6)\times SU(2)$
&7
&1
&$\s089$
&$\s013$
&$-\s083$
&$-\s032$
&$-\s012$
&$-\s012$
&$1$
\\
\rowcolor{LightGray}
$9$
&$SU(8)\times SU(2)$
&10
&0
&$1$
&$\s014$
&$-4$
&$-\s032$
&$-\s012$
&$-\s012$
&$1$
\\[.4mm]
\bottomrule
\end{tabular}
\end{center}
 \vskip-.4cm
 \caption{\label{tsolutions} Parameter and anomalous dimensions for a selection of benchmark models with superconformal fixed points and  a low number of field multiplicities, including Leigh-Strassler type models (8 and 9).}
\end{table}

The constraints  \eq{Bs}  imply that maximal values for  anomalous dimensions \eq{grange} cannot arise if gauge group factors are as small as $SU(2)$ or $SU(3)$. For larger gauge groups, however, we can find models realising maximal anomalous dimensions. 
An example  is given by an $SU(6)\times SU(2)$ gauge theory with $N_F=7$ (model 7) which has a  near-maximal  imbalance parameter $(I=\s079$). The model is located at the $\eps=-\s032$ boundary of the conformal window and leads to an asymptotically safe fixed point with  non-perturbatively large
chiral field anomalous dimensions,
\beq
\gamma_\psi =-\012\,,\quad\gamma_\Psi=-\012\,,\quad\gamma_\chi=1\,.\eeq
We conclude that the benchmark models are low-field-multiplicity realisations of asymptotic safety with phase diagrams as in Fig.~\ref{pSchematic}. Most of them are non-perturbative with large $|\eps|$ and large anomalous dimensions, whose features are not captured reliably by a few leading orders of perturbation theory (Fig.~\ref{pComparison2}). With increasing sizes of gauge group factors  $(N_1,N_2)$, many more solutions $(N_F,N_Q)$ for the constraint \eq{Bs} arise, and a fair fraction of these lead to an asymptotically safe fixed point within the UV conformal window (Fig.~\ref{pConformalWindow}).

Finally, we discuss two   benchmark  models with $\eps=-\s032$  and maximal anomalous dimensions. These strongly-coupled models are of the Leigh-Strassler type and narrowly outside the UV conformal window. The first one is model 8 with $SU(6)\times SU(2)$ gauge symmetry and  $N_F=7$, which differs from model 7 only in that $N_Q=1$ instead of  $N_Q=0$,
thus leading to a larger imbalance ($I=\s089$). 
In consequence, the  ultraviolet fixed point of model 7 degenerates into  a line of  fixed points due to $\beta_2|_*$ vanishing identically at the gauge-Yukawa fixed point. The  exactly marginal  coupling $\alpha_2$ becomes a free parameter and extends the  fixed point into an  infrared attractive line. The same phenomenon occurs for model 9, which has an $SU(8)\times SU(2)$ gauge symmetry with $N_F=10$ and imbalance $I=1$, situated at the lower-left corner of Fig.~\ref{pConformalWindow}. 
Although  interesting in their own right, the models 8 and 9 no longer serve the purpose of ultraviolet fixed points. However, we stress that 
the method of $a$-maximisation has been key in identifying  fixed points with exactly marginal directions. \step\step

\section{Conclusions} 

We have shown, as a proof of principle,  that interacting UV fixed points exist in strongly coupled quantum field theories
including away from    Veneziano (large-$N$) limits. Fixed point anomalous dimensions of matter fields  can grow large, taking values in the entire range  dictated by unitarity \eq{grange}.
Thereby, previously-found perturbative fixed points   \cite{Bond:2017suy} extend naturally into a non-perturbative conformal window  
(Fig.~\ref{pComparison1}).
  Interestingly, a novel range of  
  fixed points   has become available, 
with no perturbative counterparts  in a Veneziano limit  (Fig.~\ref{pComparison2}).  We  thus conclude that the non-perturbative ``phase space" of fixed points is large, and substantially larger then the  one accessible at weak coupling. 
 Further, all conformal fixed points  (Fig.~\ref{pConformalWindow}) are in accord with the $a$-theorem, bounds on central charges, Seiberg duality, and unitarity.  We also found a manifold  of Leigh-Strassler  models  
arising at  the sign-flip-controlling boundary of the conformal window,  where theories  display a line of IR fixed points generated by an exactly marginal gauge interaction (Fig.~\ref{pSchematicLine}).
It is understood that similar UV conformal windows,  bounded by  Leigh-Strassler manifolds, arise naturally  in other semi-simple gauge theories  coupled to matter.

From the viewpoint of particle physics and model building,
it is  quite promising that fixed points persist   at low field multiplicities (Tab.~\ref{tsolutions}),  including  with Standard Model-like gauge groups. 
Next natural  questions to ask  are whether  fixed points also arise in MSSM extensions  \cite{Hiller:2022hgt} or in  non-supersymmetric setting at strong coupling.  
From the viewpoint of conformal field theory, it will   be interesting to classify asymptotically safe SUSY theories more systematically, and to extract CFT data or structure coefficients directly from the RG fixed points.
 \step\step\step

\centerline{\bf 
 Acknowledgements}
The work of DL is supported by the Science and Technology Research Council (STFC) under the Consolidated Grant [ST/T00102X/1] .

\bibliographystyle{mystyle}
\bibliography{bib_MSSM,bib_DFL,bib_Susy}

 \end{document}